\documentclass{appolb}
\usepackage{amsmath,amsfonts,amssymb}
\usepackage{relsize}
\usepackage[dvips]{graphicx}

\begin{document}
\title{A new approach to radial spectrum of hadrons in bottom-up holographic QCD\thanks{Presented at Excited QCD 2020.}}
\author{S. S. Afonin\address{Saint Petersburg State University, 7/9 Universitetskaya nab., St.Petersburg, 199034, Russia}}
\maketitle

\begin{abstract}
Within the AdS/CFT correspondence, for description of $\mathcal{N}=4$ super Yang-Mills theory in four dimensions
one needs not only low-energy supergravity on AdS$_5$ but also the whole infinite tower of massive
Kaluza-Klein (KK) states on AdS$_5\times$S$_5$ which appear after the KK-compactification on
five-dimensional sphere. The latter aspect is usually ignored in phenomenological AdS/QCD models.
The emerging massive 5D fields on AdS$_5$ are dual to higher-dimensional
operators in 4D gauge theory, with masses being known polynomial functions of canonical dimensions
of these operators. Motivated by this observation,
we propose to calculate the spectrum of radially excited hadrons in bottom-up holographic QCD models 
as spectrum of zero KK modes of massive 5D fields dual to higher dimensional operators in QCD.
A relevant physical motivation is suggested.
The radial states with growing masses are then enumerated
by growing dimensions of interpolating QCD operators.
We tested the proposal in the Soft Wall and Hard
Wall holographic models in the sector of light mesons. The spectrum of Soft Wall
model turns out to be unchanged in the new description. But in
the Hard Wall model, our approach is shown to
lead to a much better phenomenological spectrum of vector radial excitations
than the traditional description.
\end{abstract}


\section{Introduction}

In the holographic approach to QCD, the radially excited hadrons are described as Kaluza-Klein (KK) modes
along the fifth dimension of 5D Anti-de Sitter (AdS) space. Such a description looks questionable from the physical point of view ---
hadrons are highly complicated dynamical objects in QCD while the KK-excitations are rather
simple states arising from extra dimension. The dramatic difference between the KK-like
and QCD-like states is discussed in detail in Ref.~\cite{csaki}. The main point consists
in observation that the former are deeply bound states sensitive to short distance
interactions and at collisions producing events with mostly spherical shapes while the
latter are extended states sensitive to large distance interactions and producing
characteristic jets. The underlying reason is that the latter are defined at small
't Hooft coupling $\lambda$ while the former at large $\lambda$ where the existence
of holographic duality can be motivated. The theories at small and large $\lambda$
turn out to be qualitatively different.

We propose a possible way out from theoretical
difficulties with the KK modes in the holographic hadron spectroscopy. Our proposal
consists in exploiting the higher dimensional QCD operators instead of higher
KK modes --- one should consider an infinite number of operators with growing
canonical dimension $\Delta$ for each set of quantum numbers, find the normalizable
solutions of equation of motion in a holographic model for the corresponding dual
5D field for each $\Delta$ and keep only solution with the smallest eigenvalue.
As a result, each solution will be in one-to-one correspondence with some QCD
operators. This approach looks different from the standard bottom-up AdS/QCD models
in which an infinite number of normalizable solutions (KK modes) corresponds to
only one QCD operator --- usually to an operator of leading twist 2~\cite{son2}.
We will demonstrate, however, that in the case of Soft Wall (SW) model the arising spectrum
is essentially the same as in the standard approach. In the Hard Wall (HW) model, the spectrum
becomes different and rises much slower with excitation number qualitatively
approaching the rise of experimental radial trajectories.

\section{The Soft Wall model}

For demonstration of our main idea we will
use the simplest Abelian version of the SW model~\cite{son2} defined
by the 5D action (partly different realizations of this model
are proposed in Ref.~\cite{Afonin2009})
\begin{equation}
\label{1}
S=c^2\int d^4\!x\,dz\sqrt{g}\,e^{-az^2}\left(
-\frac{1}{4}F_{MN}F^{MN}+\frac12m_5^2V_MV^M\right),
\end{equation}
where $g=|\text{det}g_{MN}|$, $F_{MN}=\partial_M V_N-\partial_N
V_M$, $M,N=0,1,2,3,4$, $c$ is a normalization constant for the
vector field $V_M$, and the background space represents the Poincar\'{e} patch
of the AdS$_5$ space with the metric
\begin{equation}
\label{2}
g_{MN}dx^Mdx^N=\frac{R^2}{z^2}(\eta_{\mu\nu}dx^{\mu}dx^{\nu}-dz^2),\qquad z>0.
\end{equation}
Here $\eta_{\mu\nu}=\text{diag}(1,-1,-1,-1)$, $R$ denotes the radius of AdS$_5$ space,
and $z$ is the holographic coordinate which is usually interpreted as the inverse energy scale.
At each fixed $z$ one has the metric of flat 4D Minkowski space.
According to the standard prescriptions of AdS/CFT correspondence~\cite{witten,gub}
the 5D mass $m_5$ is determined by the behavior of $p$-form fields near the UV boundary $z=0$,
\begin{equation}
\label{3}
m_5^2R^2=(\Delta-p)(\Delta+p-4),
\end{equation}
where $\Delta$ means the scaling dimension of 4D operator dual to the corresponding 5D field on the UV boundary.
We consider the vector case, thus $p=1$ and $m_5^2R^2=(\Delta-1)(\Delta-3)$.
The minimal value of dimension for vector operator in QCD is $\Delta=3$ that, according to~\eqref{3},
corresponds to massless 5D vector fields which are usually considered in the SW models.
But in general QCD operators interpolating vector mesons can have higher canonical dimensions, in particular,
the "descendants" preserving the chiral and Lorentz properties will
have dimensions~\cite{Afonin2019}
\begin{equation}
\label{3b}
\Delta=3+2k,\qquad k=0,1,2,\dots.
\end{equation}

The 4D mass spectrum of KK modes can be found, as usual, from the equation of
motion accepting the 4D plane-wave ansatz $V_M(x_\mu,z)=e^{ipx}v(z)\epsilon_\mu$
with the on-shell, $p^2=m^2$, and transverse, $p^\mu\epsilon_\mu=0$, conditions.
In addition, we will imply the condition $V_z=0$ for the physical
components of 5D fields. For massless vector fields, this is equivalent
to the standard choice of axial gauge due to emerging gauge invariance~\cite{son2}.
The ensuing from action~\eqref{1} equation of motion is
\begin{equation}
\label{4}
\partial_z\left(\frac{e^{-az^2}}{z}\partial_z v_n\right)=\left(\frac{m_5^2R^2}{z^3}-\frac{m_n^2}{z}\right)e^{-az^2}v_n.
\end{equation}
The particle-like excitations correspond to normalizable solutions of Sturm-Liuville equation~\eqref{4}.
It is known that they form an infinite discrete set $v_n(z)$.
The given property becomes more transparent after the substitution
\begin{equation}
\label{5}
v_n=z^{1/2}e^{az^2/2}\psi_n,
\end{equation}
which transforms the Eq.~\eqref{4} into a form of one-dimensional Schr\"{o}dinger equation
\begin{equation}
\label{6}
-\partial_z^2\psi_n+\left(a^2z^2+\frac{1+m_5^2R^2-1/4}{z^2}\right)\psi_n=m_n^2\psi_n.
\end{equation}
The mass spectrum of the model is given by the eigenvalues of Eq.~\eqref{6},
\begin{equation}
\label{8}
m_n^2=2|a|\left(2n+1+\sqrt{1+m_5^2R^2}\right),\qquad n=0,1,2,\dots.
\end{equation}
Using Eq.~\eqref{3} for $p=1$ the spectrum can be rewritten as
\begin{equation}
\label{9}
m_n^2=2|a|\left(2n+\Delta-1\right).
\end{equation}
Substituting~\eqref{3b} into this spectrum we arrive at
a remarkable relation,
\begin{equation}
\label{10}
m_n^2=4|a|\left(n+k+1\right),\qquad n,k=0,1,2,\dots,
\end{equation}
which demonstrates that within the standard SW holographic model,
the description of radial spectrum of vector mesons as zero KK modes of 5D fields dual
to higher dimensional operators is essentially the same as by KK modes stemming from a 5D field
dual to the operator of lowest dimension (the usual vector current) --- the numbers $n$ and $k$ can be interchanged.
This means that we are allowed to keep only the lightest KK mode,
$n=0$, and interpret the spectrum as arising from coupling the higher radially excited states to
higher dimensional operators constructed from the quark and gluon fields in QCD.
It is not difficult to show that the same property holds for other integer spins~\cite{Afonin2019}.

The normalized eigenfunctions corresponding to spectrum~\eqref{10} are,
\begin{equation}
\label{26b}
\phi^{(k)}_n=\sqrt{\frac{2n!}{(1+2k+n)!}}\,e^{-|a|z^2}\left(|a|z^2\right)^{1+k}L_n^{1+2k}\left(|a|z^2\right),
\end{equation}
where $L_n^\alpha(x)$ are associated Laguerre polynomials. Here the numbers $n$ and $k$ are not interchangeable
--- only the large $z$ asymptotics depends on the sum $n+k$ (because $L_n^\alpha(x)\sim x^n$ at large $x$).
The number of zeros, however, is controlled by $n$ (as the polynomial $L_n^\alpha(x)$ has $n$ zeros). By setting $n=0$, i.e. by keeping
the zero KK mode only, we choose the wave function without zeros in holographic coordinate. This wave function is the least
"entangled" with the 5th holographic dimension and thereby is the least sensitive to deviations from the AdS structure. 
The zero KK mode looks thus the most reliable in the phenomenological holographic approaches.

\section{The Hard Wall model}

Let us apply our method to HW holographic model~\cite{son1} in the case of gauge arbitrary spin fields.
Actually this analysis in the HW model was carried out for fields
dual to twist-2 operators in Ref.~\cite{katz}, we will just generalize it to arbitrary twists. The resulting
equation of motion is tantamount to making the substitution $\phi^{(J)}=z^{4-J-\Delta}\tilde{\phi}^{(J)}$ (the field
$\tilde{\phi}^{(J)}$ becomes constant at the UV boundary $z\rightarrow0$) in the corresponding equation of motion
for arbitrary spin in Ref.~\cite{katz} and drop the 5D mass term in the axial gauge. We get
\begin{equation}
\label{32}
-\partial_z\left(z^{1-2(\Delta-2)}\partial_z\tilde{\phi}^{(J)}_n\right)=m_n^2z^{1-2(\Delta-2)}\tilde{\phi}^{(J)}_n.
\end{equation}
The normalizable solution satisfying $\tilde{\phi}^{(J)}(0)=0$ is
\begin{equation}
\label{33}
\tilde{\phi}^{(J)}\sim z^{\Delta-2}J_{\Delta-2}(mz).
\end{equation}
The Dirichlet boundary condition leads to the following equation for discrete spectrum,
\begin{equation}
\label{34}
J_{\Delta-3}(m_nz_m)=0,
\end{equation}
where the property of Bessel functions $\partial_x \left(x^\alpha J_\alpha\right)=x^\alpha J_{\alpha-1}$
was exploited. Setting $\Delta=J+2$ in Eq.~\eqref{34}, we obtain the equation of Ref.~\cite{katz} for spectrum
of higher spin mesons interpolated by twist-2 operators. For arbitrary twist, we get the same pattern of degeneracy
between higher spin and radial excitations as in the SW model. The sequence of
roots of Eq.~\eqref{34} for $z_m^{-1}\approx323$~MeV and $m_0=776$~MeV is $m_nz_m\approx\{2.4,3.8,5.1,6.4,7.6,\dots\}$.
It leads to the mass spectrum
\begin{equation}
\label{35}
m_{n}\approx\{776,1234,1653,2056,2452,\dots\}.
\end{equation}
In Ref.~\cite{katz}, the spectrum~\eqref{35} was obtained for the higher spin excitations; within our approach, it coincides
with the spectrum of radially excited spin-1 $\rho$-mesons. The spectrum~\eqref{35} predicts five $\rho$-mesons below 2.5~GeV
as expected in the phenomenology~\cite{pdg} while the standard HW model does only two~\cite{son1}.

It is curious to note that the spectrum~\eqref{35} interpolates with a good precision the experimental positions for clusters of light
non-strange meson resonances arising from approximate mass degeneracies between radial and spin (or orbital) excitations~\cite{deg}.

\section{Concluding discussions}

The approach proposed in the present work may be interpreted as a simple tool establishing a direct relation between
certain gauge invariant QCD operator (or a set of) and a hadron resonance with a definite mass. 
In reality, each QCD operator has couplings to all hadron states with quantum numbers corresponding to this operator.
Indeed, calculating the two-point correlators via the standard AdS/CFT prescriptions we will get the infinite
number of pole terms corresponding to higher KK modes and which provide "meromorphization". But in our interpretation,
the accuracy of holographic models is not enough to "resolve" them as real excited hadrons, the SW model becomes an
exception (at least, until modifications of metric and background are introduced). A somewhat similar situation
takes place in lattice simulations: In order to "resolve" radially excited hadron states one invokes higher dimensional
QCD operators~\cite{Dudek}. 

In the holographic dual picture, something should hamper the excitation of higher KK modes along the fifth dimension in AdS$_5$ space. 
A possible mechanism has recently been suggested in Ref.~\cite{Fichet:2019hkg}: 
Gravity in AdS dresses free propagators on the quantum level leading to universal exponential suppression of propagators in the infrared region.
In other words, the AdS space turns out to be opaque to propagation in deep infrared region. This naturally locks the excitation
of any higher KK mode.

From the viewpoint of original AdS/CFT correspondence, the KK-states are not gone. 
We recall that in order to describe the $\mathcal{N}=4$ super Yang-Mills theory in four dimensions, 
one must use not only low-energy supergravity on AdS$_5$ but also the whole infinite tower of massive 
KK-states on AdS$_5\times$S$_5$~\cite{witten} which appear after the KK-compactification on 
five-dimensional sphere. It becomes natural to suggest that the 5D fields dual to higher-dimensional 
operators in 4D gauge theory correspond to these KK states from S$_5$.

\end{document}